\documentclass[12pt]{article}

\usepackage{graphicx}

\newcommand{\ket}[1]{|\, #1 \rangle}
\newcommand{\bra}[1]{\langle #1 \,|}
\newcommand{\udt}[3]{#1^{#2}_{\phantom{#2}#3}}
\newcommand{\dut}[3]{#1_{#2}^{\phantom{#2}#3}}

\begin{document}
\begin{center}
{\Large\bf Spin decoherence by spacetime curvature}
\vskip .6 cm
Hiroaki Terashima and Masahito Ueda
\vskip .4 cm
{\it Department of Physics, Tokyo Institute of Technology,\\
Tokyo 152-8551, Japan and \\
CREST, Japan Science and Technology Corporation (JST),\\
Saitama 332-0012, Japan}
\vskip .6 cm
\end{center}

\begin{abstract}
A decoherence mechanism caused
by spacetime curvature is discussed.
The spin state of a particle is shown to decohere
if only the particle moves in a curved spacetime.
In particular, when a particle is near
the event horizon of a black hole,
an extremely rapid spin decoherence occurs for
an observer who is static in a Killing time,
however slow the particle's motion is.
\end{abstract}

\begin{flushleft}
{\scriptsize
}
\end{flushleft}

\section{Introduction}
In quantum information processing~\cite{NieChu00},
the spin of a spin-$1/2$ particle
is often utilized as a qubit
regardless of the momentum state of the particle.
However, the spin and momentum
are not, in general, separable
in the relativistic regime,
since the Lorentz transformation (LT) entangles
them via the Wigner rotation~\cite{Wigner39}.
Among recent work~\cite{PeScTe02,AlsMil02,%
TerUed02,GinAda02,LiDu02,CzaWil03}
on entanglement in relativity, 
Peres et al.\ showed that
the spin entropy of a spin-$1/2$ particle
is not invariant under the LT
unless the particle is
in the momentum eigenstate~\cite{PeScTe02}.
This means that even if
the spin is in a pure state in one frame of reference,
it may not be so in another
unless the entanglement with the momentum is
taken into account.
Thus, the spin alone cannot be used as a qubit
in quantum information processing that involves
relativistically moving observers.

In this paper,
we address the question of how the spin of a particle
moving in a gravitational field decoheres
due to the effects of general relativity.
In general relativity, a gravitational field
is represented by a spacetime curvature,
which entails a breakdown of
the global rotational symmetry.
The spin in general relativity can therefore be
defined only locally by invoking
the rotational symmetry
of the local inertial frame.
As a consequence of this local definition,
the motion of the particle is accompanied by
a continuous succession of LTs.
We shall show that this effect,
which is caused by spacetime curvature,
gives rise to a spin entropy production
that is unique to general relativity.
This means that
even if the state of the spin is pure at one spacetime point,
it, in general, becomes mixed at an other spacetime point.
As an illuminating example,
we shall show that the spin entropy
of a circularly moving particle changes extremely rapidly
near the event horizon of a black hole
in a local inertial frame
that is static with respect to
a Killing time at each point.

It is here in order to remark
two important distinctions between the present
general-relativistic problem
and the special-relativistic one discussed
in Refs.~\cite{PeScTe02,AlsMil02,TerUed02,GinAda02,LiDu02}.
First, while spin entropy is not invariant under the LT
in special relativity~\cite{PeScTe02},
it \emph{is} invariant under
the general coordinate transformation (GCT)
in general relativity.
The spin is defined relative to
the local inertial frame,
which is physically determined by
the spacetime distribution of matter,
whereas the GCT is an artificial relabeling
of spacetime points, and therefore does not affect
the local inertial frame, leaving the spin state invariant.
The transformation that changes the spin entropy
is a local LT of the local inertial frame.
Second, while in special relativity
spin entropy is altered by changing
the inertial frame,
in general relativity
spin entropy can change by
a \emph{mere translation} of the particle
even though both the general coordinate system
and the local inertial frame are fixed at each point;
here the spin entropy can be generated
because local inertial frames at different points
are, in general, different, as shown below.
An entropy production of this type cannot
be found in other general-relativistic problems
that do not involve the spin degrees
of freedom~\cite{vanRud03,KokYur03}.

This paper is organized as follows.
Section~\ref{sec:formulation} explains our formulation of
a relativistic spin in a curved spacetime.
Section~\ref{sec:decoherence} shows that
a spin decoherence is caused by spacetime curvature.
Section~\ref{sec:example} considers an example
in the Schwarzschild spacetime.
Section~\ref{sec:conclude} summarizes our results.

\section{\label{sec:formulation}Formulation}
A gravitational field in general relativity
is described by a curved spacetime
with metric $g_{\mu\nu}(x)$.
To discuss the spin of a particle
in the curved spacetime,
we introduce a local inertial frame at each point.
The coordinate transformation
from a general coordinate system $x^\mu$
to the local inertial frame $x^a$ at each point
can be carried out using a vierbein (or a tetrad)
$\dut{e}{a}{\mu}(x)$ defined by~\cite{Nakaha90}
\begin{equation}
 \dut{e}{a}{\mu}(x) \, \dut{e}{b}{\nu}(x)
  \,g_{\mu\nu}(x)=\eta_{ab},
\label{eq:vierbein}
\end{equation}
where $\eta_{ab}=\mathrm{diag}(-1,1,1,1)$
is the Minkowski metric.
Here and henceforth, it is assumed that
Latin and Greek letters run over
the four inertial-coordinate labels $0,1,2,3$
and the four general-coordinate labels, respectively,
and that repeated indices are to be summed.
The vierbein transforms a tensor
in a general coordinate system $x^\mu$
into that in a local inertial frame $x^a$, and vice versa.
For example, momentum $p^\mu(x)$
in the general coordinate system can be transformed
into that in the local inertial frame at $x^\mu$
via the relation $p^a(x)=\udt{e}{a}{\mu}(x)\, p^\mu(x)$.
The inverse of the vierbein $\udt{e}{a}{\mu}(x)$
is defined by
\begin{equation}
 \udt{e}{a}{\mu}(x)\,\dut{e}{a}{\nu}(x)=\dut{\delta}{\mu}{\nu},\qquad
 \udt{e}{a}{\mu}(x)\,\dut{e}{b}{\mu}(x)=\udt{\delta}{a}{b}.
\end{equation}

The choice of the local inertial frame is not unique,
since the inertial frame remains inertial under the LT.
The choice of the vierbein therefore has
the same degree of freedom known as the local LT.
It is this degree of freedom that
transforms the spin of a particle.
Namely, a spin-$1/2$ particle in the curved spacetime
is defined as a particle whose one-particle states
furnish the spin-$1/2$ representation of the local LT,
not of the GCT.
Note that the Dirac field in the curved spacetime
is spinor under the local LT,
whereas it is scalar under the GCT.
Usually, the definition of a particle is not unique
in quantum field theory in curved spacetime~\cite{BirDav82},
because we cannot uniquely choose the time coordinate
to define the positive energy.
However, in the present formulation,
our particle is specified by the choice of the vierbein,
since $\dut{e}{0}{\mu}(x)$ generates
a preferred global time coordinate from
the local inertial time coordinate (the $0$-axis).

Consider a wave packet of
a spin-$1/2$ particle with mass $m$ in the curved spacetime.
We assume that the spacetime scale of the wave packet
is sufficiently small compared with
that over which the curvature changes.
This assumption allows us to
refer to the wave packet as a ``particle.''
Let us suppose that
the centroid of the wave packet is located
at point $x^\mu$ and
is moving with four-velocity $dx^\mu/d\tau=u^\mu(x)$,
which is normalized as $u^\mu(x)u_\mu(x)=-c^2$;
this motion is not necessarily geodesic
in the presence of an external force.
The momentum of the centroid is then given by 
$q^a(x)=\udt{e}{a}{\mu}(x)\left[m u^\mu(x)\right]$
in the local inertial frame at the point $x^\mu$.
Moreover, using this local inertial frame,
we can describe the wave packet as in special relativity.
Namely, the momentum eigenstate $\ket{p^a,\sigma}$ of the particle
is labeled by the four-momentum
$p^a=(\sqrt{|\vec{p}|^2+m^2c^2},\vec{p})$ and
by the $z$-component $\sigma$ ($=\uparrow$, $\downarrow$)
of the spin~\cite{Weinbe95}.
The wave packet is then expressed
as a linear combination of $\ket{p^a,\sigma}$,
\begin{equation}
 \ket{\psi}=\sum_\sigma \int d^3\vec{p}\;N(p^a)
 \,C(p^a,\sigma) \,\ket{p^a,\sigma},
\end{equation}
where
\begin{equation}
d^3\vec{p}\;N(p^a)\equiv
d^3\vec{p}\;\frac{mc}{\sqrt{|\vec{p}|^2+m^2c^2}}
\end{equation}
is a Lorentz-invariant volume element.
From the normalization condition
\begin{equation}
\langle p'^a,\sigma'\,|\,p^a,\sigma \rangle=\frac{1}{N(p^a)}\,
\delta^3(\vec{p'}-\vec{p})\,\delta_{\sigma'\sigma},
\end{equation}
we find that the coefficient $C(p^a,\sigma)$ satisfies
\begin{equation}
  \sum_\sigma \int d^3\vec{p}\;N(p^a)
   |C(p^a,\sigma)|^2=1.
\end{equation}
Taking the trace of the density matrix
$\rho=|\psi\rangle\langle\psi|$ over the momentum,
we obtain the reduced density matrix
for the spin,
\begin{eqnarray}
  \rho_\mathrm{r}(\sigma';\sigma)
  &=& \int d^3\vec{p}\;N(p^a)\,
      \bra{p^a,\sigma'}\rho\ket{p^a,\sigma} \nonumber \\
  &=& \int d^3\vec{p}\;N(p^a)\,
   C(p^a,\sigma')\,C^\ast(p^a,\sigma).
\end{eqnarray}
The spin entropy of the wave packet is
the von Neumann entropy of
this reduced density matrix:
\begin{equation}
  S=-\mathrm{Tr}\left[\rho_\mathrm{r}(\sigma';\sigma)
     \log_2\rho_\mathrm{r}(\sigma';\sigma)\right].
\end{equation}

It is important to note that
the spin index $\sigma$ denotes not Dirac spin
but Wigner one~\cite{Wigner39},
which is based on the Poincar\'{e} symmetry.
Using the Pauli matrices $\vec{\sigma}$,
Wigner spin operator for the wave packet is given by~\cite{Terno02}
\begin{equation}
\hat{\vec{S}}=\frac{1}{2}\sum_{\alpha,\beta}\vec{\sigma}_{\alpha\beta}
  \int d^3\vec{p}\;N(p^a) \,\ket{p^a,\alpha} \bra{p^a,\beta},
\end{equation}
while Dirac spin operator is given by
\begin{equation}
\hat{\vec{\Sigma}}=
 \int d^3x : \hat{\psi}^\dagger_m(x) \left[
 \frac{1}{2} \left(\begin{array}{cc}
    \vec{\sigma} & 0 \\
    0 & \vec{\sigma}
  \end{array}\right) \right]_{mn}
 \hat{\psi}_n(x):,
\end{equation}
where $\hat{\psi}_m(x)$ $(m=1,2,3,4)$ is a Dirac spinor field and
the colons $::$ mean the normal ordering.
As is well known, Dirac spin is not a conserved quantity,
since $\hat{\vec{\Sigma}}$ does not commute with the Hamiltonian.
In contrast, Wigner spin is a conserved quantity,
since $\hat{\vec{S}}$ does commute with the Hamiltonian.
Because Wigner spin is defined using the particle's rest frame,
it is analogous to the non-relativistic spin.
Thus, with Wigner spin, we can discuss a spin observable
even in a relativistic regime.

\section{\label{sec:decoherence}Decoherence}
After an infinitesimal proper time $d\tau$,
the centroid of the wave packet moves to
a new point $x'^\mu =x^\mu+u^\mu(x)d\tau$,
and the wave packet is then described
by the local inertial frame at the new point $x'^\mu$.
In the new local inertial frame,
the momentum of the centroid changes to
$q^a(x')=q^a(x)+\delta q^a(x)$
due to an acceleration by the external force
\emph{and} due to a change in the local inertial frame.
The explicit form of $\delta q^a(x)$ is
given by~\cite{TerUed03b}
\begin{equation}
 \delta q^a(x)= \left[ma^a(x)+
 \udt{\chi}{a}{b}(x)\,q^b(x)\right]d\tau,
\label{eq:delq}
\end{equation}
where
\begin{equation}
 a^a(x)=\udt{e}{a}{\mu}(x)\,\left[\,u^\nu(x)
 \nabla_\nu u^\mu(x)\,\right]
\end{equation}
is the acceleration by the external force and
\begin{equation}
\udt{\chi}{a}{b}(x)
=u^\mu(x) \left[\,\dut{e}{b}{\nu}(x)\nabla_\mu
    \udt{e}{a}{\nu}(x)\,\right]
\end{equation}
is the change in the local inertial frame along $u^\mu(x)$.
Since $q^a(x)q_a(x)=-m^2c^2$ and $q^a(x)a_a(x)=0$,
the change $q^a(x)\to q^a(x)+\delta q^a(x)$
may be interpreted as a local LT
$\udt{\delta}{a}{b}+\udt{\lambda}{a}{b}(x)d\tau$, where
\begin{equation}
 \udt{\lambda}{a}{b}(x)=-\frac{1}{mc^2}\left[\,a^a(x)\,q_b(x)
 -q^a(x)\,a_b(x)\,\right]+\udt{\chi}{a}{b}(x).
\label{eq:ill}
\end{equation}
While the first term due to the acceleration
exists even in special relativity,
the second term due to the spacetime curvature
is unique to general relativity.
Note that even if the wave packet moves as straight as possible
(i.e., moves along a geodesic curve $a^a(x)=0$),
this LT may be nonzero in general relativity.
By iterating this infinitesimal transformation,
we obtain a transformation formula
for a finite proper time as a time-ordered Dyson series,
since $\udt{\lambda}{a}{b}(x)$'s at different points
do not necessarily commute.
When the centroid of the wave packet moves
along a path $x^\mu(\tau)$
from $x_i^\mu=x^\mu(\tau_i)$ to $x_f^\mu=x^\mu(\tau_f)$,
the motion of the wave packet is accompanied by a LT given by
\begin{equation}
   \udt{\Lambda}{a}{b}(x_f,x_i)=
   T\exp\left[\int^{\tau_f}_{\tau_i}
    \udt{\lambda}{a}{b}(x(\tau))\, d\tau\right],
\label{eq:ll}
\end{equation}
where $T$ is the time-ordering operator.

Since the spin entropy is not invariant under the LT,
neither is it invariant during
the motion of the wave packet.
Note that the momentum eigenstate $\ket{p^a,\sigma}$
transforms under a LT $\udt{\Lambda}{a}{b}$
as~\cite{Weinbe95,Ohnuki88}
\begin{equation}
 U(\Lambda)\, \ket{p^a,\sigma}
 =\sum_{\sigma'} D^{(1/2)}_{\sigma'\sigma}
 (W(\Lambda,p))\,\ket{\Lambda p^a,\sigma'},
\end{equation}
where $D^{(1/2)}_{\sigma'\sigma}(W(\Lambda,p))$ is
the $2\times 2$ unitary matrix that represents
the Wigner rotation $\udt{W}{a}{b}(\Lambda,p)$;
the explicit form of the Wigner rotation reads
\begin{equation}
\udt{W}{a}{b}(\Lambda,p)=
\left[L^{-1}(\Lambda p)\,\Lambda\,L(p)\right]\udt{}{a}{b},
\label{eq:wigner}
\end{equation}
where $\udt{L}{a}{b}(p)$ describes a standard LT defined by
\begin{eqnarray}
\udt{L}{0}{0}(p) &=& \gamma, \nonumber \\
\udt{L}{0}{i}(p) &=& \udt{L}{i}{0}(p)=p^i/mc,    \\
\udt{L}{i}{k}(p) &=& \delta_{ik}+
(\gamma-1)\,p^i\,p^k/|\vec{p}|^2,\nonumber
\end{eqnarray}
with $\gamma=\sqrt{|\vec{p}|^2+m^2c^2}/mc$ and $i,k=1,2,3$.
Since the Wigner rotation of the spin depends on the momentum,
the LT entangles the spin with the momentum. 
In accordance with the LT (\ref{eq:ll})
induced by the motion of the wave packet,
the reduced density matrix for the spin
in the local inertial frame at $x^\mu_f$ becomes
\begin{eqnarray}
 \rho'_\mathrm{r}(\sigma';\sigma) &=&
   \sum_{\sigma''\sigma'''}\int d^3\vec{p}\;N(p^a)
  \,C(p^a,\sigma'')\,C^\ast(p^a,\sigma''')
  \nonumber \\
  & & \qquad{}\times
  D^{(1/2)}_{\sigma'\sigma''}(W(\Lambda(x_f,x_i),p))
   \nonumber \\
  & & \qquad{}\times
  D^{(1/2)\ast}_{\sigma\sigma'''}(W(\Lambda(x_f,x_i),p)).
\end{eqnarray}
This reduced density matrix $\rho'_\mathrm{r}(\sigma';\sigma)$
at $\tau_f$, in general, represents a mixed state,
even if the initial reduced density matrix
$\rho_\mathrm{r}(\sigma';\sigma)$ represents
a pure state at $\tau_i$.
This means that the spin entropy is generated by
both gravity and acceleration during
the motion of the wave packet.

\section{\label{sec:example}Example}
As an illustrative example, we consider
the Schwarzschild spacetime~\cite{Wald84},
\begin{eqnarray}
 g_{\mu\nu}(x)dx^\mu dx^\nu&=&-f(r)c^2dt^2+\frac{1}{f(r)}dr^2
 \nonumber \\
&& +r^2(d\theta^2+\sin^2\theta d\phi^2),
\end{eqnarray}
where $f(r)=1-(r_s/r)$, with $r_s$ being the Schwarzschild radius.
At this radius,
the spacetime has the event horizon,
on which the coordinate system $(t,r,\theta,\phi)$ breaks down.
The time coordinate $t$ is known as a Killing time,
with respect to which the Schwarzschild spacetime is static.
In the Schwarzschild spacetime,
we introduce a static observer with
a static local inertial frame at each point
by choosing the vierbein (\ref{eq:vierbein}) as
\begin{eqnarray}
\dut{e}{0}{t}(x)=\frac{1}{c\sqrt{f(r)}} ,&\quad&
\dut{e}{1}{r}(x)=\sqrt{f(r)} , \nonumber \\
\dut{e}{2}{\theta}(x)=\frac{1}{r}, &\quad&
\dut{e}{3}{\phi}(x)=\frac{1}{r\sin\theta},
\label{eq:stvb}
\end{eqnarray}
with all the other components being zero.
Below, only non-zero components are shown.
Note that the inertial frame is defined
at each instant,
since the observer is accelerated to keep staying
at the given point.
Such an accelerated observer would perceive
the Hawking radiation~\cite{Hawkin75}
if the state of the quantum field is represented
in a Fock space which is defined using the Kruskal time.
However, in the present example,
the observer does not suffer from the Hawking radiation,
because we consider a Fock space
which is defined using the Killing time.
Actually, by the choice of the vierbein (\ref{eq:stvb}),
the local inertial time coordinate $x^0$
is parallel to the Killing time coordinate $t$.

Suppose that the centroid of the wave packet
is moving along a circular trajectory of
radius $r$ ($>r_s$) with a constant velocity
$rd\phi/dt\equiv v\sqrt{f(r)}$ on the equatorial plane
$\theta=\pi/2$.
The four-velocity of the centroid is then given by
\begin{equation}
 u^t(x)=\frac{\cosh\xi}{\sqrt{f(r)}}, \qquad
 u^\phi(x)=\frac{c\sinh\xi}{r},
\end{equation}
where $\xi$ is the rapidity
in the local inertial frame defined by
\begin{equation}
\tanh\xi=\frac{v}{c}.
\end{equation}
Accordingly,
in the local inertial frame at any point,
the four-momentum of the centroid becomes
$q^a(x)=(mc \cosh\xi,0,0,mc \sinh\xi)$.
We assume that the coefficient of the wave packet
at $\tau_i$ is
\begin{eqnarray}
 C(p^a,\sigma) &=&
 \frac{\delta_{\sigma,\uparrow}}{\sqrt{\pi^{1/2}wN(p^a)}}
 \exp\left[-\frac{\left(p^3-q^3(x)\right)^2}{2w^2}\right]
  \nonumber \\
 & & \qquad\qquad\qquad\qquad{}\times
   \sqrt{\,\delta(p^1)\,\delta(p^2)\,},
\end{eqnarray}
which is Gaussian in $p^3$
with the spin $z$-component $\uparrow$,
even though $p^1$ and $p^2$ have the definite value $0$.
The spin entropy of this wave packet is zero at time $\tau_i$,
since the momentum is not entangled with the spin.

Since the wave packet is not in a geodesic motion,
the first term in the infinitesimal LT (\ref{eq:ill})
is not zero in this example, giving rise to
a generalized Thomas precession~\cite{TerUed03b}
in the curved spacetime.
This effect is an origin of the spin-orbit coupling
and contributes to the spin entropy
even in the limit of Minkowski spacetime $r_s\to0$.
In contrast, the second term in Eq.~(\ref{eq:ill}),
which consists of a boost along the $1$-axis
\begin{equation}
\udt{\chi}{0}{1}(x) =
\udt{\chi}{1}{0}(x) =
-\frac{cr_s\cosh\xi}{2r^2\sqrt{f(r)}}
\end{equation}
and a rotation about the $2$-axis
\begin{equation}
\udt{\chi}{1}{3}(x) =
-\udt{\chi}{3}{1}(x) =
\frac{c\sinh\xi\sqrt{f(r)}}{r},
\end{equation}
causes the spin decoherence that is unique to general relativity.
In fact, in the limit of Minkowski spacetime,
this term is reduced to a pure rotation,
which does not change the spin entropy as shown below.

Combining these terms, we see that the infinitesimal LT
$\udt{\lambda}{a}{b}(x)$ becomes
\begin{eqnarray}
\udt{\lambda}{0}{1}(x) &=& \udt{\lambda}{1}{0}(x)=-L\tanh\xi, 
\label{eq:boo} \\
\udt{\lambda}{1}{3}(x) &=& -\udt{\lambda}{3}{1}(x)=L,
\label{eq:rot}
\end{eqnarray}
where
\begin{equation}
 L=\frac{c\cosh^2\xi\sinh\xi}{r}
   \left[1-\frac{r_s}{2rf(r)}\right]\sqrt{f(r)}.
\end{equation}
After a proper time
$\tau_p=\tau_f-\tau_i$ of the particle,
the Wigner rotation (\ref{eq:wigner})
for the finite LT (\ref{eq:ll})
is reduced to a rotation about the $2$-axis
through angle $\Omega(p^a)\,\tau_p$, where
\begin{equation}
\Omega(p^a) =
\left[1-\frac{p^3}{p^0+mc}\tanh\xi\right]L.
\end{equation}
The first and second terms in $\Omega(p^a)$
correspond to the rotation part (\ref{eq:rot})
and the boost part (\ref{eq:boo}),
respectively.
The corresponding unitary representation
can be expressed in terms of
the Pauli matrix $\sigma_y$ as
\begin{equation}
D^{(1/2)}_{\sigma'\sigma}(W(\Lambda(x_f,x_i),p))
=\exp\left(-i\frac{\sigma_y}{2}\,\Omega(p^a)\,\tau_p\right).
\end{equation}
Therefore, the reduced density matrix at $\tau_f$ becomes
\begin{equation}
 \rho'_\mathrm{r}(\sigma';\sigma)=\frac{1}{2}
  \left(\begin{array}{cc}
 1+\overline{\cos\Omega\tau_p}
 & \overline{\sin\Omega\tau_p} \\[8pt]
 \overline{\sin\Omega\tau_p} &
 1-\overline{\cos\Omega\tau_p}
  \end{array}\right),
\end{equation}
where the overline means
the average over the momentum distribution,
$\overline{X} = \int d^3\vec{p}\;N(p^a)
|C(p^a,\uparrow)|^2\, X(p^a)$.

As a result, the spin entropy generated
by the gravity and acceleration is given by
\begin{equation}
 S'=-P\log_2 P-(1-P)\log_2(1-P),
\label{eq:genent}
\end{equation}
where
\begin{equation}
 P=\frac{1}{2}\left[1-
 \left|\overline{\exp\left(i\Omega\tau_p\right)}
 \right|\right].
\end{equation}
Note that only the $p^a$-dependent term in $\Omega(p^a)$,
which arises from the boost part (\ref{eq:boo}),
contributes to the spin entropy.
It is easy to see that $0\le P\le 1/2$ and $0\le S'\le 1$,
since
\begin{equation}
0\le
\left|\overline{\exp\left(i\Omega\tau_p\right)}
\right|^2
\le
\overline{\left|\exp\left(i\Omega\tau_p\right)
\right|^2}=1.
\label{eq:prange}
\end{equation}
If the wave packet is in
the momentum eigenstate $w=0$,
spin entropy is not generated, i.e. $S'=0$,
since then the second inequality in Eq.~(\ref{eq:prange})
is saturated.

\begin{figure}
\begin{center}
\includegraphics[scale=0.8]{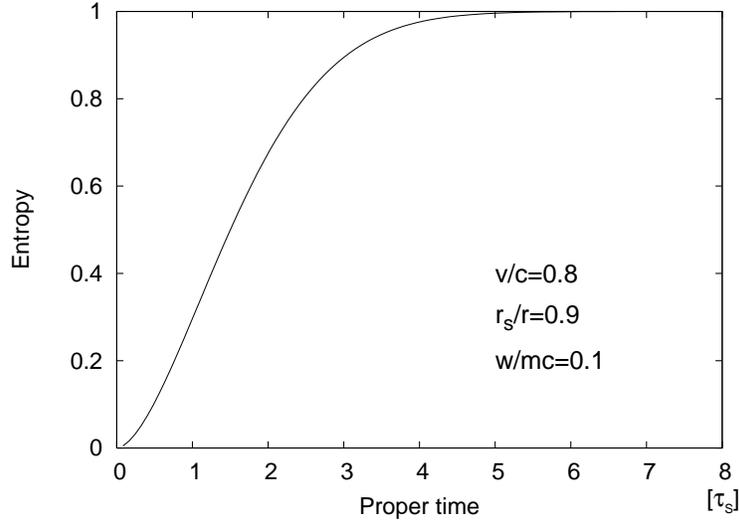}
\end{center}
\caption{\label{fig1}The generated entropy $S'$ as
a function of the proper time $\tau_p$ at $v/c (=\tanh\xi)=0.8$,
$r/r_s=0.9$, and $w/mc=0.1$.
The proper time is normalized by $\tau_s=mr_s/w$.}
\end{figure}

\begin{figure}
\begin{center}
\includegraphics[scale=0.8]{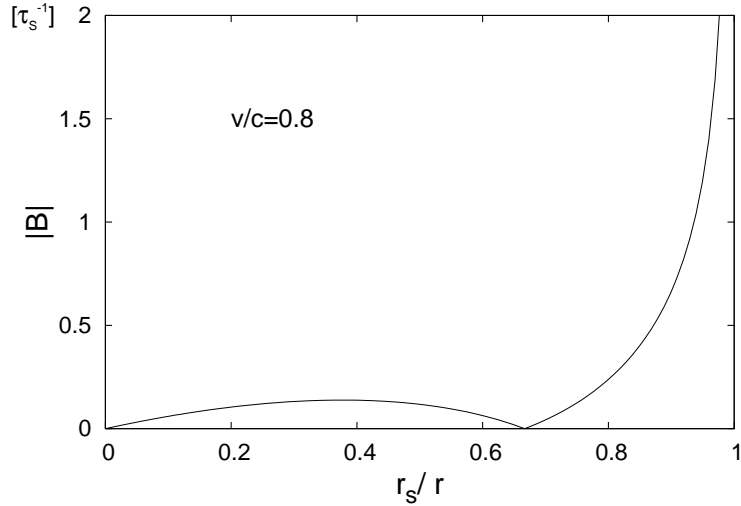}
\end{center}
\caption{\label{fig2}The inverse of the characteristic
decoherence time, $|B|$,
as a function of $r_s/r$ at $v/c=0.8$,
normalized by $\tau_s^{-1}=w/mr_s$.
$|B|=0$ at $r=\infty$ or $3r_s/2$,
whereas $|B|=\infty$ at $r=r_s$.}
\end{figure}

In the case of $w/mc\ll 1$,
expanding $\Omega(p^a)$ around $q^3(x)=mc\sinh\xi$
up to the second order in $(p^3-q^3(x))/mc$,
we obtain
\begin{equation}
 \left|\overline{\exp\left(i\Omega\tau_p\right)}
 \right| \simeq \frac{1}{(1+A^2\tau_p^2)^{1/4}}
 \exp\left[-\frac{B^2\tau_p^2}{4(1+A^2\tau_p^2)}
 \right],
\end{equation}
where
\begin{eqnarray}
  A &=& \frac{Lw^2\tanh^2\xi}{2m^2c^2}
  \left[\frac{1}{(\cosh\xi+1)^2}-\frac{1}{\cosh^2\xi}\right], \\
  B &=& \frac{Lw\tanh\xi}{mc}
  \left[\frac{1}{\cosh\xi}-\frac{1}{\cosh\xi+1}\right].
\end{eqnarray}
We thus find that
the reduced density matrix decoheres to
a mixed state $S'>0$
after the proper time $\tau_p\sim |B|^{-1}$
and becomes maximally mixed ($S'\to 1$)
in the limit of $\tau_p=\infty$,
as shown in Fig.~\ref{fig1}.
No revival of coherence occurs
as opposed to the case of an electron in an atom.
In this latter case the electron has a discrete spectrum
and no real decoherence occurs,
while in our case the momentum distributes continuously
due to a semiclassical prescription in which
the orbital motion is fixed to a given one
by an appropriate external force at a macroscopic radius.
Figure~\ref{fig2} shows the inverse of the characteristic
decoherence time, i.e. $|B|$,
as a function of $r_s/r$.
The spin decoherence is extremely rapid
near the event horizon,
because $|L|$ becomes very large as $r$ approaches $r_s$.
We emphasize that this is true even when
the wave packet moves with non-relativistic velocity,
if $r$ is sufficiently close to $r_s$ or
if $w$ is sufficiently large.
In particular, right on the event horizon,
the reduced density matrix becomes
maximally mixed in an infinitesimal proper time.
Of course, the formula (\ref{eq:genent})
no longer holds inside the event horizon $r<r_s$,
since the strong gravity makes it impossible
for the particle to move circularly.
At the spatial infinity $r\to\infty$, on the other hand,
the reduced density matrix remains pure ($S'=0$),
since both gravity and acceleration vanish there.
An interesting situation occurs
on the sphere of the radius $r=3r_s/2$,
on which the reduced density matrix does not decohere
because the LT caused by gravity is canceled
by that caused by acceleration, giving $L=0$.

In contrast to the static observer,
an observer moving with the wave packet
does not see the spin decoherence,
because $\delta q^a(x)$ in Eq.~(\ref{eq:delq})
is zero in the co-moving local inertial frame.
However, these two results do not contradict each other,
since the spin entropy is not invariant
under the LT from a static local inertial frame
to a co-moving one.

\section{\label{sec:conclude}Conclusion}
In conclusion,
we have shown that spin entropy
is generated when a spin-$1/2$ particle
moves in a gravitational field.
Even if the spin at one spacetime point
is in a pure state,
it may evolve into a mixed state
as the particle moves.
In particular,
the spin entropy of a circularly moving particle
increases very rapidly near the event horizon of
the Schwarzschild black hole.
Unless the entanglement with the momentum is
taken into account,
the spin cannot be used as a qubit
in quantum information processing
where the system is subject to
a strong gravitational field.

\section*{Acknowledgments}
H.T. was partially supported by
the Japan Society for the Promotion of Science.
This research was supported by a Grant-in-Aid
for Scientific Research (Grant No.15340129) by
the Ministry of Education, Culture, Sports,
Science and Technology of Japan.


\begin{thebibliography}{10}

\bibitem{NieChu00}
M.~A. Nielsen and I.~L. Chuang, {\em Quantum Computation and Quantum
  Information} (Cambridge University Press, Cambridge, 2000).

\bibitem{Wigner39}
E.~P. Wigner, Ann. Math. {\bf 40},  149  (1939).

\bibitem{PeScTe02}
A. Peres, P.~F. Scudo, and D.~R. Terno, Phys. Rev. Lett. {\bf 88},  230402
  (2002).

\bibitem{AlsMil02}
P.~M. Alsing and G.~J. Milburn, Quantum Inf. Comput. {\bf 2},  487  (2002).

\bibitem{TerUed02}
H. Terashima and M. Ueda, Quantum Inf. Comput. {\bf 3},  224  (2003);
Int. J. Quantum Inf. {\bf 1},  93  (2003).

\bibitem{GinAda02}
R.~M. Gingrich and C. Adami, Phys. Rev. Lett. {\bf 89},  270402  (2002);
R.~M. Gingrich, A.~J. Bergou, and C. Adami, Phys. Rev. A {\bf 68},  042102
  (2003).

\bibitem{LiDu02}
H. Li and J. Du, Phys. Rev. A {\bf 68},  022108  (2003).

\bibitem{CzaWil03}
M. Czachor and M. Wilczewski, Phys. Rev. A {\bf 68},  010302(R)  (2003).

\bibitem{vanRud03}
S.~J. van Enk and T. Rudolph, Quantum Inf. Comput. {\bf 3},  423  (2003);
P.~M. Alsing and G.~J. Milburn, Phys. Rev. Lett. {\bf 91},  180404  (2003).

\bibitem{KokYur03}
P. Kok and U. Yurtsever, Phys. Rev. D {\bf 68},  085006  (2003).

\bibitem{Nakaha90}
M. Nakahara, {\em Geometry, Topology and Physics} (Institute of Physics
  Publishing, Bristol, 1990).

\bibitem{BirDav82}
N.~D. Birrell and P.~C.~W. Davies, {\em Quantum Fields in Curved Space}
  (Cambridge University Press, Cambridge, 1982).

\bibitem{Weinbe95}
S. Weinberg, {\em The Quantum Theory of Fields} (Cambridge University Press,
  Cambridge, 1995).

\bibitem{Terno02}
D.~R. Terno, Phys. Rev. A {\bf 67},  014102  (2003).

\bibitem{TerUed03b}
H. Terashima and M. Ueda, Phys. Rev. A {\bf 69},  032113  (2004).

\bibitem{Ohnuki88}
Y. Ohnuki, {\em Unitary Representations of the Poincar{\'e} Group and
  Relativistic Wave Equations} (World Scientific, Singapore, 1988).

\bibitem{Wald84}
R.~M. Wald, {\em General Relativity} (University of Chicago Press, Chicago,
  1984).

\bibitem{Hawkin75}
S.~W. Hawking, Commun. Math. Phys. {\bf 43},  199  (1975).

\end{thebibliography}

\end{document}